[Tapez ici]

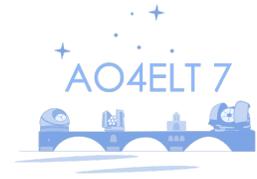

# V(WF)²S: Very Wide Field WaveFront Sensor for GLAO

Olivier Lai[*a], Mark Chun[b], Stefan Kuiper[c], Niek Doelman[c,d], Marcel Carbillet[a], Mamadou N'Diaye[a], Frantz Martinache[a], Lyu Abe[a], Jean-Pierre Rivet[a], Dirk Schmidt[e]

[a] Université Côte d'Azur, Observatoire de la Côte d'Azur, CNRS, Laboratoire Lagrange, Bd de l'Observatoire, CS 34229, 06304 Nice Cedex 4, France; [b]IfA, University of Hawaii, 640 North A'ohoku Place, Hilo, HI 96720 USA; [c]TNO Technical Sciences, Stieltjesweg 1, 2628CK Delt, The Netherlands; [d]Leiden University, Niels Bohrweg, 2, 2333CA Leiden; [e]National Solar Observatory, 3665 Discovery Drive, Boulder, CO 80303, USA.

## ABSTRACT

Adaptive optics is an advanced technique developed for and mostly used on large telescopes. It turns out to be challenging for smaller telescopes (0.5~2m) due to the small isoplanatic angle, small subapertures and high correction speeds needed at visible wavelengths, requiring bright stars for guiding, severely limiting the sky coverage. Natural guide star (NGS) SCAO (Single Conjugate Adaptive Optics) is ideal for planetary objects but remains limited for general purpose observing. The approach we consider is a compromise between image quality gain and sky coverage: our proposition is that it is better to partially improve the image quality anywhere in the sky than to be limited by diffraction around a few thousand bright stars. We therefore suggest a new solution based on multiple fundamental AO concepts brought together to enable a whole new field of application: The principle is based on a rotating Foucault test, like the first AO concept proposed by Horace Babcock in 1953, on the Ground Layer Adaptive Optics (GLAO), proposed by Rigaut and Tokovinin in the early 2000s, and on the idea of Layer-oriented MCAO and the pupil-plane wavefront analysis by Roberto Ragazzoni. We propose to combine these techniques to use all the light available in a large field to measure the ground layer turbulence and enable the high angular resolution imaging of regions of the sky (e.g., nebulas, galaxies) inaccessible to traditional SCAO systems.

The motivation to develop compact and robust AO system for small telescopes is two-fold: On the one hand, schools and universities often have access to small telescopes as part of their education programs. Also, researchers in countries with fewer resources could also benefit from well-engineered and reliable adaptive optics system on smaller telescopes for research and education purposes. On the other hand, amateur astronomers and enthusiasts want improved image quality for visual observation and astrophotography. Implementing readily accessible adaptive optics in astronomy clubs would also likely have a significant impact on citizen science.

## 1. INTRODUCTION

Adaptive optics is a technique used to compensate for atmospheric turbulence induced image distortion and degradation at the focus of astronomical telescopes [1]. Most major observatories are equipped with unique systems using state of the art components, working at near infrared wavelengths; it turns out to be difficult to adapt this technique to smaller, i.e., 0.5 to 1.5m class, telescopes for e.g., astrophotography. This is due to the fact the subapertures of the wavefront sensors are necessarily small, the sampling frequency must be high and the isoplanatic angle is small in the visible domain; this translates into a low limiting magnitude (between 5 and 7) allowing the observation of a few thousands of bright stars; Natural guide star (NGS) SCAO (Single Conjugate Adaptive Optics) is ideal for planetary objects but remains limited for general purpose observing. It is also impractical to deploy laser guide stars (LGS) for small telescopes for obvious safety and operational reasons. Therefore, to increase the sky coverage and potential observable sources, we propose the use of GLAO (Ground Layer Adaptive Optics) which enables wide fields to be corrected, although not at the diffraction limit [2, 3]. To this end we have developed a "Very Wide Field WaveFront Sensor", V(WF)$^2$S, which uses the light of all the stars in an area as large as many tens of arcminutes to perform the wavefront measurement in a single pupil plane [4].

---

[*] olivier.lai@oca.eu

The V(WF)$^2$S is based on the principle of optical differentiation and uses an active mask, such as a Spatial Light Modulator (SLM) or Digital Light Processor (DLP), to perform a linear optical operation in the focal plane which allows to transform a function of the phase (in this case, the derivative) into an intensity function, which can be measured. After experimenting with pyramids, sawtooth and other asymmetric gradient masks, we found that the most efficient modulation is a rotating Foucault knife edge test, as originally proposed by H. Babcock in his foundational paper of 1953: "The possibility of compensating astronomical seeing" [5], which is in fact very similar to a pyramid wavefront sensor with each of the four pupil images taken sequentially instead of simultaneously.

## 1.1 Motivation

To enable large sky coverage, we first recognize that we cannot correct the entire volume of turbulence above the telescope or in a direction where there is no source to illuminate the wavefront perturbation we are trying to correct. The next best thing we can do is to correct the turbulence which is common to the entire field we are trying to capture. This is achieved by GLAO, whereby a single deformable mirror or lens, conjugated to the telescope pupil is used to correct only the ground layer turbulence [3]. This is obtained by averaging wavefront measurements in multiple directions. The corrected wavefront will be limited by the residuals of the free atmosphere turbulence, so resulting images will in general not be diffraction limited but will however show a substantial improvement in FWHM [4].

The V(WF)$^2$S concept truly stands on the shoulders of giants, and the only innovation is the use of an addressable focal plane mask to perform the optical differentiation operations sequentially. But this dynamic modulation introduces other advantages such as the use of a *single detector* and *no moving parts*, the ability to adapt the wavefront sensor to the scene (or even the type of object, e.g., stars, planets, nebulas and galaxies, but also solar granulation, ocular retinas or live cells *in vivo*), the possibility of using a binary (as opposed to greyscale) masks (if useful/required), as well as blocking out the sky background where no useful sources are present.

The concept we propose opens whole new fields of applications on smaller telescopes. This potentially makes adaptive optics accessible to whole new communities, especially considering that there are many more telescopes in the 1-2m class than in the 8-10 m class (it turns out telescopes follow Zipf's Law [6]):

- On the one hand, schools and universities often have access to small telescopes as part of their education programs. Furthermore, researchers in countries with fewer resources would also benefit from well-engineered and robust adaptive optics for smaller telescopes, by improving their performance and exposure to advanced optics.

- On the other hand, amateur astronomers and enthusiasts would be keen for improved image quality for visual observation and astrophotography. Implementing readily accessible adaptive optics in astronomy clubs would also likely have a significant impact on citizen science.

Our goal is to develop a closed loop wide-field AO system for a reasonable budget (≲k€20), but we note in passing that there are potentially many other applications, notably in the fields of AO-assisted microscopy and ophthalmology, as the V(WF)$^2$S can be adapted to any extended incoherent source for wavefront sensing measurements. Furthermore, layer-oriented MCAO could be easily implemented with a single camera and no moving parts, simply by programming the active mask in the focal plane to sequentially select wide and narrow constellations to perform the required tomography. Finally, the V(WF)$^2$S could also be used on larger telescopes of the 8-10m class or even ELTs in principle; most such systems will be equipped with laser guide stars which makes NGS wavefront sensing less essential. Furthermore, the Lagrange Invariant becomes very large and requires gigantic optics to transmit wide fields. Using smaller fields increases the probability of large magnitude differences in the guide sources, which reduces the homogeneity of the correction. Nonetheless an infrared version of the W(WF)$^2$S could offer substantial sky coverage even with a limited (2~5') field of view.

## 2. EXPERIMENTAL SETUP

### 2.1 Optical averaging demonstrator

We have assembled a simple prototype to demonstrate the principle of operation of the V(WF)$^2$S. The experimental protocol was developed to demonstrate optical averaging, by confirming that the flux weighted average of wavefront measurements from many individual sources would be the same as the simultaneous wavefront measurement on all sources. As shown in Figure 1, the images of multiple stars are focused onto an SLM by lens L2, which defines the optical system's pupil; the pupil is reimaged on the detector by lens L3 and images are recorded as the Foucault knife edge rotates around each star (see Section 2.3).

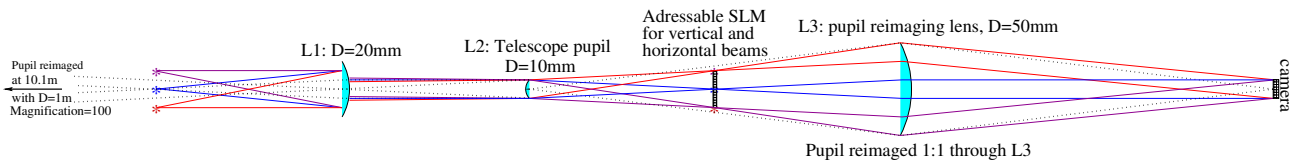

Figure 1. The virtual image of multiple sources is formed at infinity (on the left) by lens L1, as seen by the telescope pupil at lens L2. These sources are reimaged at the focal plane of lens L2 where an addressable intensity mask is placed. Lens L3 reimages the pupil L2 on the detector. The modulation of intensity in the focal plane can be adapted to the scene, but in the case of point sources, a rotating Foucault knife edge test on each star is the simplest and most efficient form of optical differentiation.

For this demonstrator, we are currently using a HoloEye LC 2012 SLM as the addressable intensity mask in a focal plane to generate an optical differentiation pattern on the extended source of interest. The SLM is sufficient for our purposes, but being monochromatic, working in polarised light and being relatively slow (60Hz) makes it unsuitable for sky operation. We will eventually upgrade to a Texas Instrument DLP - e.g. DLP670S, when speed and resolution become more critical; however, this will also require syncing the modulator with the acquisition of the pupil plane camera.

### 2.2 Assembly

The four stars are created from a single laser diode at 635nm and 1x4 fiber splitter. This has the unfortunate consequence that the sources are coherent amongst themselves and generate a fringe pattern in the pupil. The sources are placed sufficiently far apart that the fringe pattern is at very high frequency and under-sampled on the camera, but fringes could still be seen on final pupil images due to aliasing.

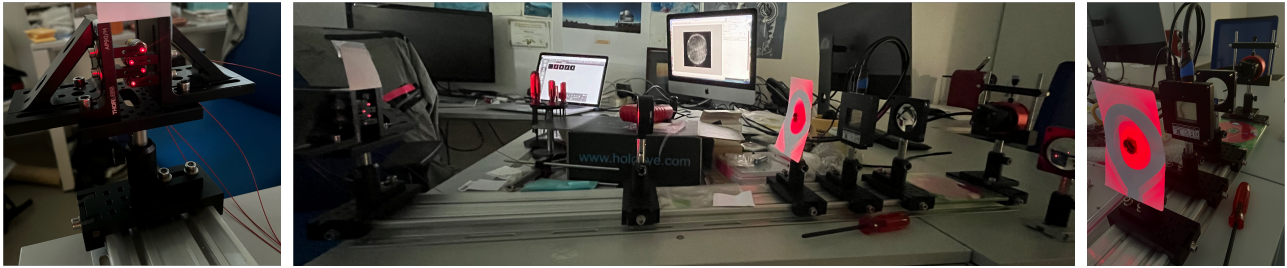

Figure 2. The four sources are generated by single mode fibers fed by a single laser diode via a 1x4 splitter (left). The pupil plane is located at the illuminated piece of paper, followed by the SLM and lens L3 and finally the camera on the right. The images of the stars are (faintly) visible on the SLM in the picture on the right.

The SLM pixel pitch is 36µm, and when it is transparent, it acts like a diffraction grating, producing replicas of the pupil on the camera. In the original design, the pupil was 10mm in diameter, and the FWHM of unresolved sources was ~6.4µm; being smaller than the pixel pitch meant that the different orders of the diffraction of the pupil were overlapping. The pupil had to be stopped down to 1.8mm, producing PSFs with a FWHM of 35.3µm such that the pupil images did not overlap anymore. The importance of the diffraction due to the grating in a focal plane only became apparent during the experimental phase and will be a critical part for the design of the DLP based system.

## 2.3 Mask design

In the demonstrator, we generated binary masks using PowerPoint to generate .png images which were uploaded to the SLM. Without a focal plane viewing option, this process would be impossible, so we introduced a 75mm focal length lens between L3 and the camera to reimage the focal plane on the detector. We positioned each "D" shape (and its rotations and mirrors) on each star by hand (which was still a tedious process). In a system with an imager, a correspondence between the focal plane pixels of the science camera and of the focal plane modulator will have to be established (a one-time daytime calibration using a flat-field and a grid pattern on the intensity mask) and a high pass filtered image of the science field can then be used to generate the binary mask on the fly.

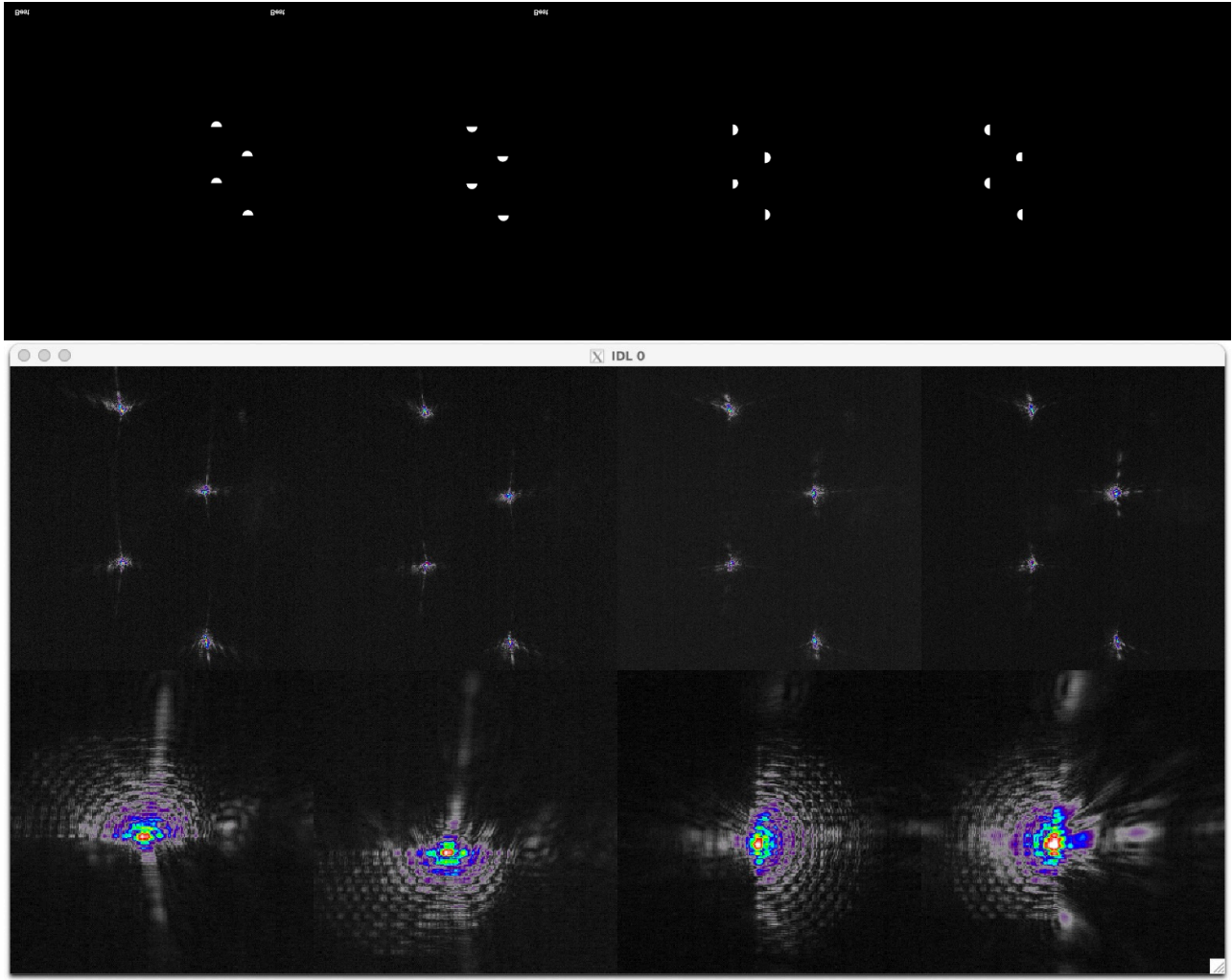

Figure 3: Binary masks (top) for each of the four rotations of the "D" mask. Below, focal plane images for the same masks, showing some field/static aberrations and diffraction.

Two images are obtained for the horizontal gradient (x-slope) and two images for the vertical gradient (y-slope). With four stars simultaneously, we obtain four images of the pupil, which we combine as normalised contrast:

$$slope_x = \frac{ima_1 - ima_2}{ima_1 + ima_2}$$

$$slope_y = \frac{ima_3 - ima_4}{ima_3 + ima_4}$$

With four sequential stars, we obtain 16 images, used to generate four individual x-slopes and four y-slopes, which are flux weighted averaged.

## 2.4 Results – no phase screen

Results obtained without any additional phase (i.e., with the system's own aberrations) are shown in Figure 4. These results were obtained by playing a slide show on the SLM and asynchronously recording images of the pupil on the camera. Hundreds of images were averaged for each position of the mask to reduce the impact of turbulence within the office.

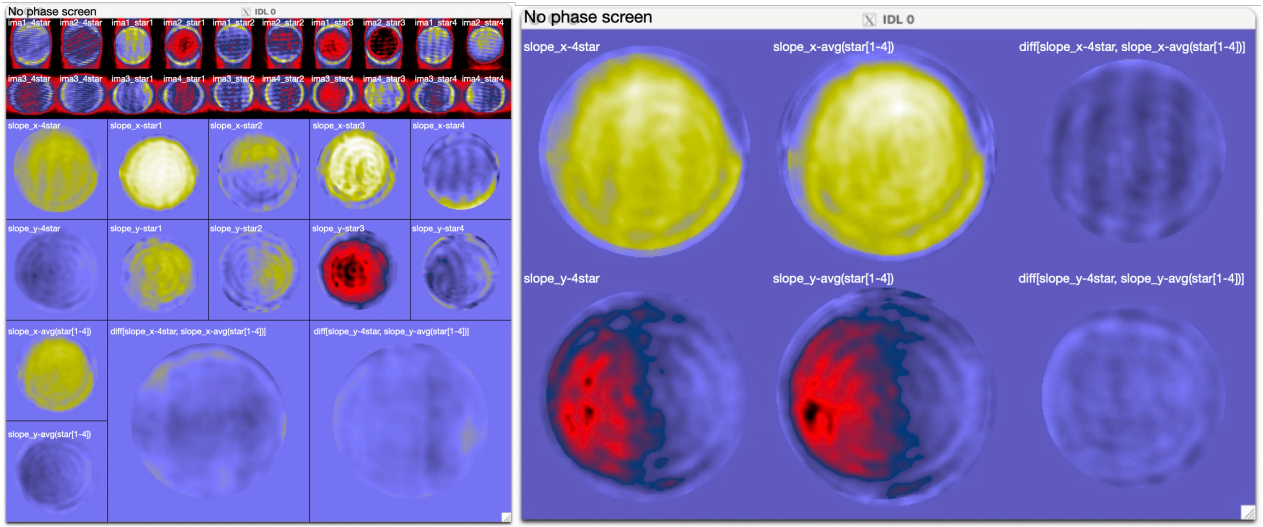

Figure 4: Results of the comparison between the simultaneous measurement of four stars and the flux weighted average of individual star measurement. See text for left panel; on the right panel, the x-slopes are shown on top, y-slopes at the bottom, and from left to right: 4 stars simultaneously; flux weighted of individual star measurement; difference (same scale). Fringes are visible due to the aliasing of the fringe pattern from the 4 sources in the pupil.

The left panel of figure 4 shows the pupil images in the top two rows (x-slope at top, y-slope at bottom), and from left two right, the two images for the simultaneous measurement on all stars and the eight images (four stars x 2 images per star) of the individual star measurements. The next two rows show the normalized contrast for the simultaneous measurement and the 4 individual stars, x-slope at top, y-slope at bottom; the bottom row shows the flux weighted average of the 4 stars (left) and the x-slope (middle) and y-slope (right) differences. These are shown a stretched scale on the right panel. Astigmatism is present, as seen by the gradient on the y-slope. Fringes are visible in the difference images; these are due to aliasing of the fringe pattern of the four coherent sources in the pupil.

## 2.5 Results with Phase screen

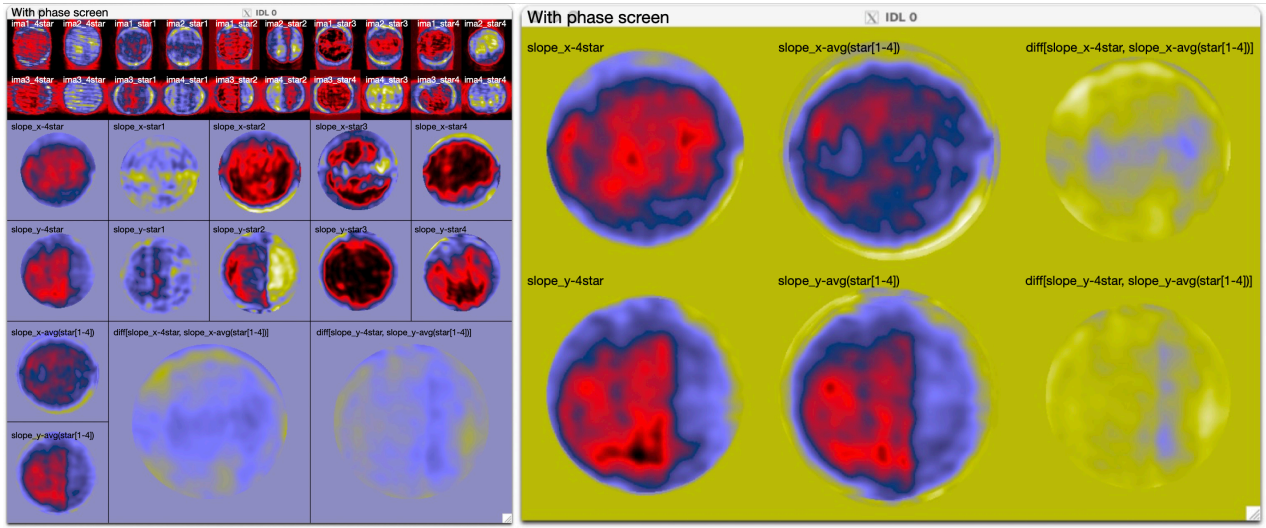

Figure 5: same as Figure 4 but with a phase screen in the beam. See text for details.

A phase screen (a CD cover) was introduced in the optical beam between L1 and L2 and measurements were carried out again. These are shown in Figure 5, with the same layout as Figure 4.

As was the case without a phase screen, the difference images show that the flux-weighted average of the measurements is indeed very similar to the optical averaging taking place in the simultaneous measurement of the four sources. It is particularly interesting to note that the beam on source 2 has a sharp step on the y-slope, but it is correctly measured in the average measurement. There are still small differences, and we posit that these might be due to:
- Office turbulence,
- Non-linear behavior of Foucault knife edge on individual sources with strong signal
- Pupils not perfectly aligned (simple optical set-up using plano-convex singlets!)

Coherent sources were a big drawback of experiment, with persistent fringing (through aliasing) but will not be an issue on stellar sources. It is nonetheless a **successful proof of concept experiment to demonstrate multi-object wavefront sensing and optical averaging**: We obtained a satisfactory comparison of a random wavefront measured with 4 sources individually to flux-weighted average of measurement with 4 sources simultaneously.

## 2.6 Extended source

Although not displayed on the poster, we obtained another significant result using the V(WF)$^2$S on an extended source: We replaced the four individual sources by a single object, "too large to be a moon", which we illuminated with white light, shown on Figure 6.

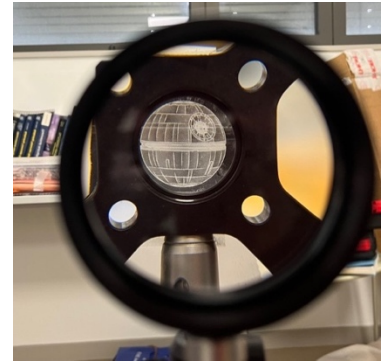

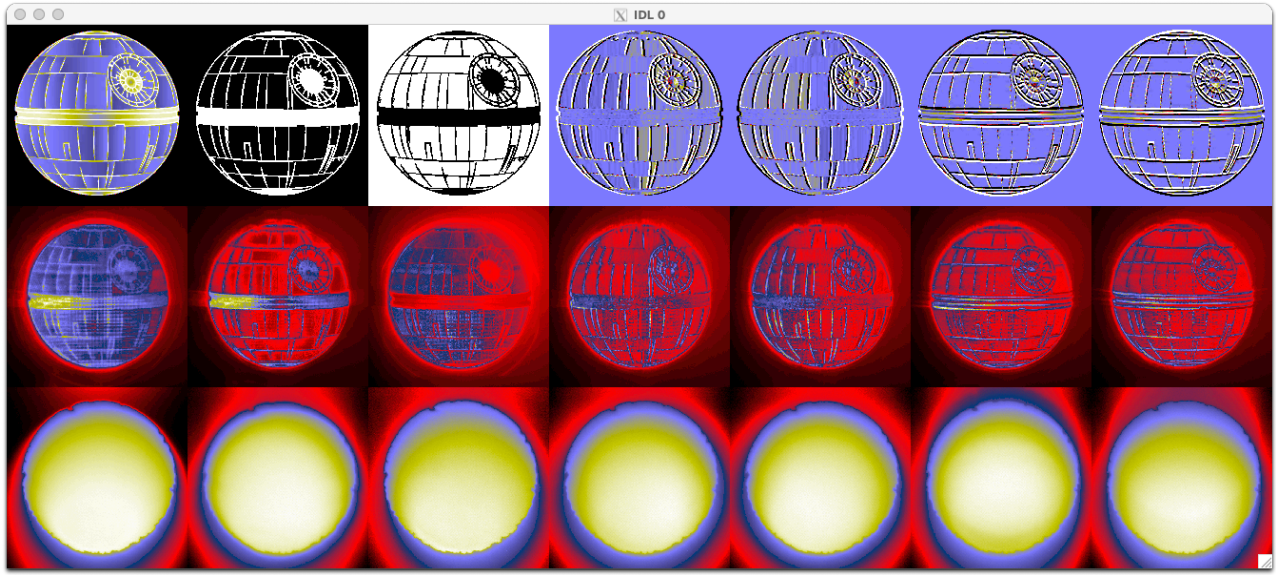

Figure 6: extended source used at the input focus of the demonstrator (right), "too big to be a moon". The right panel shows the mask applied on the SLM (top), the focal plane recorded in imaging mode (middle) and the pupil plane (bottom). From left to right: the input object, a positive mask (increasing the contrast of sharp lines), a negative mask (masking sharp line), the two x-slope masks (the x-derivative of the input mask, stretched to saturate slightly) and the two y-slope masks.

Using such an extended source and white light illumination (although still using a narrow band filter centered on 635nm in the camera) removed all the fringing that was visible when using the coherent sources. Results of the difference between a measurement without and with a phase screen are shown on the right panel of Figure 7. We did not have an alternative way to measure the phase introduced by the phase screen to determine whether this difference was an accurate estimate. In fact, determining the linearity and dynamic range of this sensor remains to be carried out.

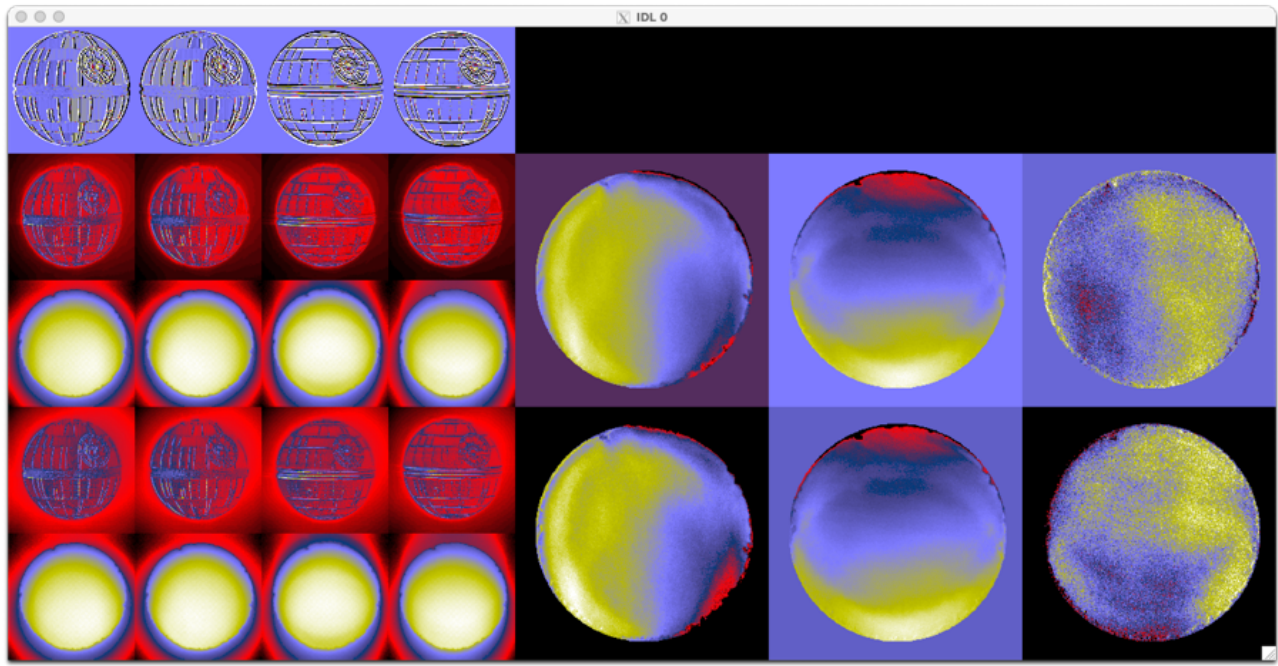

Figure 7: top row, left shows the mas applied on the SLM. The next two rows show the focal and pupil plane without a phase screen, and the bottom two rows with the phase screen. The right panel show the x-slope (top) and y-slope (bottom) for the normalized contrast without the phase screen, with the phase screen and their difference (same scale).

## 3. FUTURE DEVELOPMENTS

### 3.1 Closed loop demonstrator using a DLP

We propose the schematic design shown in Figure 8 for visible GLAO for 1m class telescope, using a single camera for multi-object wavefront sensing and an adaptive lens for wavefront correction; the transmissive elements make the optical layout simple, but the DLP requires an off-axis reflection. The size of the WFS field is envisaged to be 10'~20' but will eventually be determined by the size of L3 (although a clever optical design with appropriate pixel size on the camera can probably reduce its size – or increase the size of the wavefront sensing field).

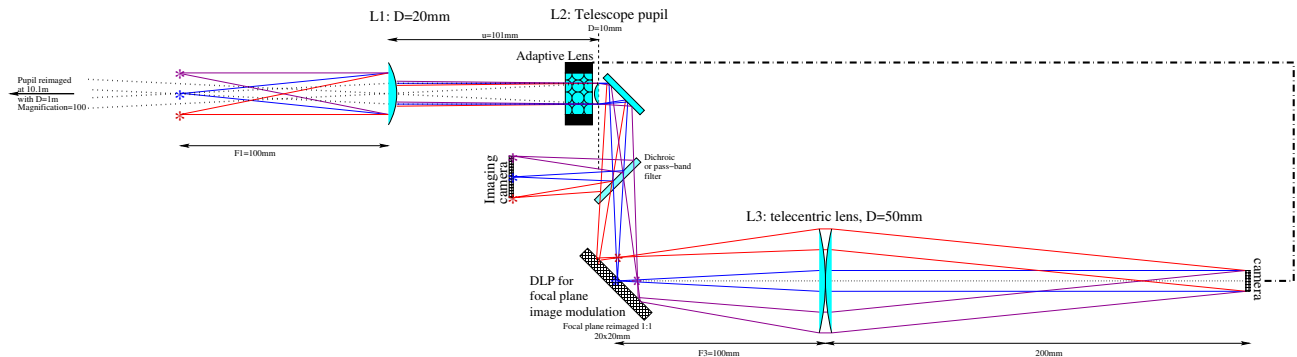

Figure 8: Schematic layout for a closed loop wide field AO system using a DLP for focal plane modulation and an adaptive lens for the wavefront correction and a single detector, synched with the DLP.

We envisage setting up this demonstrator first in the laboratory:

- The first step will be to implement and test a DLP as a focal plane mask. Due to its reflective nature, the optical layout will need to be 2D (as opposed to the current 1D rail); we aim to minimize the reflection angle on the DLP to reduce the defocus at the edges of the field, and may consider a "Batman" type-design as proposed by F. Zamkotsian [7]. At this stage, we will need to develop a controller scheme to be able to synch the DLP and the camera.

- We will then introduce a deformable lens at the entrance pupil of the system. Different models are available, such as the one from Dynamic Optics in Italy (https://www.dynamic-optics.it/deformable-lenses/) or Phaseform GmbH in Germany (https://www.phaseform.com/). Besides closed loop demonstration, this step will enable the measurement of the linearity and dynamic range of the V(WF)$^2$S as we will be able to introduce known and calibrated phase aberrations. This stage will require the deployment of a real time controller capable of communicating with the DLP, camera and deformable lens.

- Finally, we will move the eventually optimized layout on a stand-alone optical bench in order to deploy it at the Cassegrain focus of the C2PU 1m telescope at Calern observatory.

Possible future plans also include implementing a focal plane mask in the wavefront sensor of the AOC (Adaptive Optics at Calern) [8] system for its planetary mode (which has to be able to perform wavefront measurements on Jupiter, currently an under-sampled Shack-Hartmann sensor with a 1' FoV per sub-aperture).

### 3.2 Expected performance.

Simulations of the V(WF)$^2$S wavefront sensor in closed loop GLAO correction show a potential improvement of a factor 2 in FWHM and a factor 2 in peak intensity (Strehl ratio) under average atmospheric turbulence conditions. As shown on Figure 9, this is a substantial gain for astrophotography, where we have used a *Hubble Space Telescope* image of Messier 22 and degraded it to good seeing (0.6", left) and GLAO corrected seeing (0.3", middle; the original HST image has a resolution of 0.07").

Note that it would be relatively easy to extend the capability from GLAO to layer oriented MCAO, by using two sets of masks (one for the narrow and one for the wide constellations) sequentially and simply adding another adaptive lens at the appropriate conjugation. Corrected MCAO fields are usually smaller than GLAO fields, but if the diffraction limit of a 1m telescope in the visible (0.1", a factor 3 improvement in the middle bottom panel of Figure 9!) is achievable on a sufficient number of asterisms, this may be of interest too.

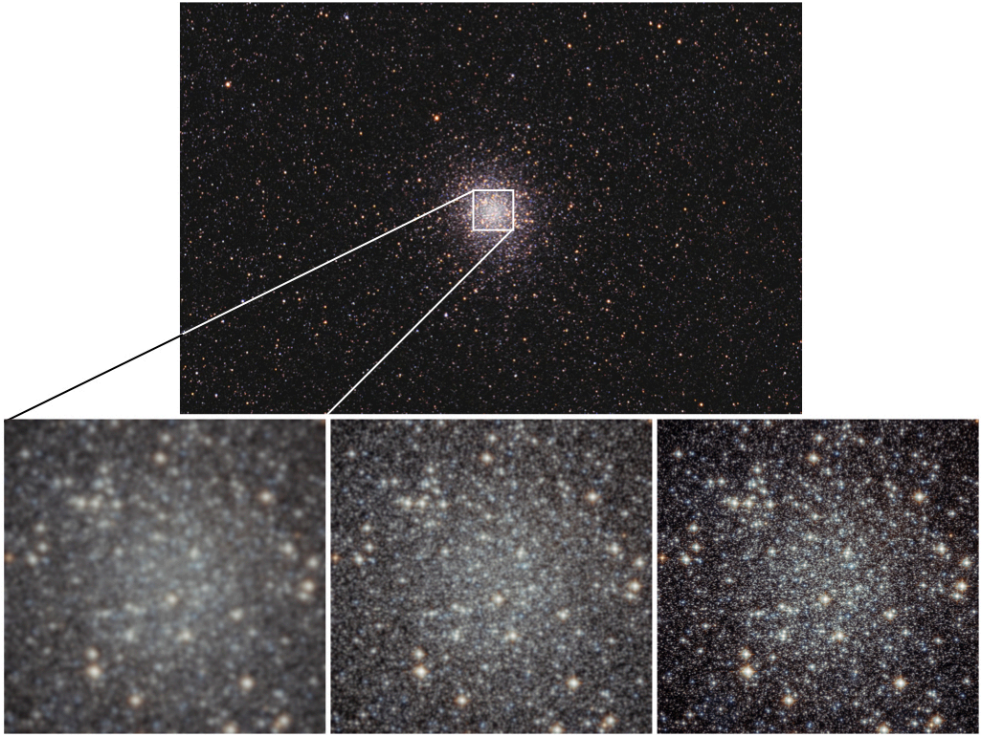

Figure 9, top: typical image of M22 obtained by an amateur telescope. Bottom row, left: seeing limited; right: Hubble Space Telescope image; middle: typical expected GLAO performance on 1m telescope.

## REFERENCES

[1] F. Roddier, "Adaptive Optics in Astronomy", Edited by François Roddier, pp. 419. ISBN 0521612144. Cambridge, UK: Cambridge University Press, November 2004.
[2] F. Rigaut, "Ground conjugate wide field adaptive optics for the ELTs", *Proc. Topical meeting "Beyond conventional adaptive optics"*, ESO Conference and Workshop Proceedings **58**, ISBN 3923524617, p.11 (2002)
[3] A. Tokovinin, "Seeing improvement with ground layer adaptive optics", *Pub. Astron. Soc. Pacific*, **116**, 941-951 (2004)
[4] Roberto Ragazzoni, "Pupil plane wavefront sensing with an oscillating prism", Journal of Modern Optics, **43**, 289-293, DOI: 10.1080/09500349608232742 (1996)
[5] H. Babcock, "The possibility of compensating astronomical seeing", *Pub. Astron. Soc. Pacific*, **65**, 229 (1953).
[6] O. Lai, S. Kuiper, N. Doelman, M.Chun, D. Schmidt, F. Martinache, M. Carbillet, M. N'Diaye, J.-P. River, "Making adaptive optics available to all : a concept for 1m class telescopes", *Proc SPIE Adaptive Optics Systems VIII*, **12185**,1218575 (2022)
[7] F. Zamkotsian, P. Lanzoni, N. Tchoubaklina et al., "Batman @ TNG : instrument integration and performance", *Proc. SPIE Ground based and airborne instrumentation for Astronomy VII* **10702**, 107025P (2018).
[8] B. Buralli, O. Lai, M. Carbillet, L. Abe, L. Abe, F.-X. Schmider, J. Dejonghe, F. Martinache, E. Aristidi, E. Cottalorda, Y. Bresson, J.-P. Rivet, D. Vernet, F. Vakili, "Numerical modelling of the planetary adaptive optics mode of AOC, the adaptive optics project at Calern Observatory", *Proc. SPIE Adaptive Optics Systems VIII* **12185**, 121858R (2022)

*This work was supported by the Action Spécifique Haute Résolution Angulaire (ASHRA) of CNRS/INSU co-funded by CNES.*